\documentclass[preprint2]{emulateapj}

\usepackage{multirow}
\shorttitle{Fe K lines in \objectname{NGC 253}}
\shortauthors{Mitsuishi et al.}

\begin{document}

\title{Fe K line complex in the nuclear region of \objectname{NGC 253}}

\author{Ikuyuki Mitsuishi, Noriko Y. Yamasaki and Yoh Takei}
\affil{Institute of Space and Astronautical Science, Japan Aerospace Exploration Agency (ISAS/JAXA),\\
3-1-1 Yoshinodai, Chuo-ku
Sagamihara, Kanagawa, 252-5210, Japan}

\begin{abstract}
A bright, nearby 
edge-on starburst galaxy \objectname{NGC 253} was studied using
the {\it Suzaku}, {\it XMM} and {\it Chandra} X-ray observatories.
We detected with {\it Suzaku} and {\it XMM} 
complex line structure of Fe K, which is resolved into 
three lines (\ion{Fe}{1} at 6.4 keV, \ion{Fe}{25} at 6.7 keV and \ion{Fe}{26} at 7.0 keV)
around the center of \objectname{NGC 253}. 
Especially, the \ion{Fe}{1} and \ion{Fe}{26} lines are the first clear detections,  
with a significance of $>$99.99 \% and 99.89 \% estimated  by a Monte Carlo procedure.
Imaging spectroscopy with {\it Chandra} revealed that 
the emission is distributed in $\sim$60 arcsec$^2$ region around the nucleus,
which suggests that the source is not only the buried AGN. 
The flux of highly ionized Fe lines can be explained by the 
accumulation of 10--1000 supernova remnants that are the 
result of high starforming activity, while the \ion{Fe}{1}
line flux is consistent with the fluorescent line
emission expected with the molecular clouds in the region.
\end{abstract}

\keywords{galaxies: individual (\objectname{NGC 253}) --- galaxies: ISM --- galaxies: starburst --- galaxies: star formation --- X-rays: galaxies}

\section{Introduction}

X-ray emission from the central region of a starburst galaxy is known to be complex 
because of blending of various kinds of point sources and diffuse emission. 
Although point source emission typically dominates the hard band, 
in some starburst galaxies including \objectname{NGC 253}, 
the emission line from He-like \ion{Fe}{25} is observed,
suggesting the existence of a high temperature plasma
\citep[e.g.][]{ngc253-nuclear-xmm,ngc6240-xmm,m82-xmm}.
Supernova remnants (SNRs), associated with starburst activity,
are considered to be a candidate of the origin.
Moreover, two of the galaxies (NGC6240 and M82) exhibit
\ion{Fe}{1} fluorescent line
\citep{ngc6240-xmm,m82-fe-k-complex}.
The coexistence of highly ionized and neutral Fe lines suggests 
the coexistence of hot and cold gases.
Thus, this complex Fe K line structure is informative on
the origin of the emission and 
the gas structure of starburst galaxies.
The origin of the neutral line is, however, still poorly known.

\objectname{NGC 253} is a  nearby bright edge-on starburst galaxy 
with the star formation rate of 1.4--9.5 M$_{\odot}$ yr$^{-1}$ \citep{ngc253-sfrs}.
Its large apparent diameter makes it
a suitable target to study the spacial distribution of emission
features.
{\it Chandra} suggested that the X-ray emission in the nuclear region 
is composed of a low-temperature thermal plasma and a heavily absorbed hard component 
that is expressed by a power law plus emission from a photoionized plasma 
emitting \ion{Fe}{4}-\ion{Fe}{15} and \ion{Fe}{25} \citep{chandra-nuclear-agn-suggestion}.
The hard component is interpreted as a buried AGN \citep{ngc253-agn}.
The spectrum around the nuclear region could be modeled with 
three-temperature thin thermal plasmas including a high 
($\sim$6~keV) temperature plasma responsible for
the \ion{Fe}{25} line \citep{ngc253-nuclear-xmm}. 

In this Letter, we report on the study of the nuclear region of
\objectname{NGC 253} using 
{\it Suzaku} and {\it Chandra}, with a particular focus on the Fe line complex. 
{\it Suzaku} has a good energy resolution to resolve the Fe line complex  
with a low background, 
while a sub-arcsecond spatial resolution of {\it Chandra} enables us to argue 
 spatial distribution.
We also show results 
by {\it XMM-Newton} as a complementary analysis.  
Throughout this paper, we adopt the distance to \objectname{NGC 253} of 3.4 Mpc \citep{ngc253-distance} 
corresponding to 16 pc arcsec$^{-1}$.
We assume the solar abundance tabulated by 
\citet{solar-abundance-table-anders}.
HEAsoft version 6.11, SAS 10.0.0, CIAO 4.3 and XSPEC 12.7.0 were used for
data reduction and spectral analysis.
For all fitting processes, we adopted C statistics and 
binned spectra so that each bin contains $>10$ counts after a background subtraction.
We quote an error as a 90 \% significance level of a single parameter in all tables, determined by Monte Carlo simulations and 
most of errors are consistent with those obtained by $\Delta$C-statistic of 2.7.

\section{Observations \& Data Reduction}

\begin{figure*}[htb!]
\begin{center}
  \includegraphics[width=5.4cm]{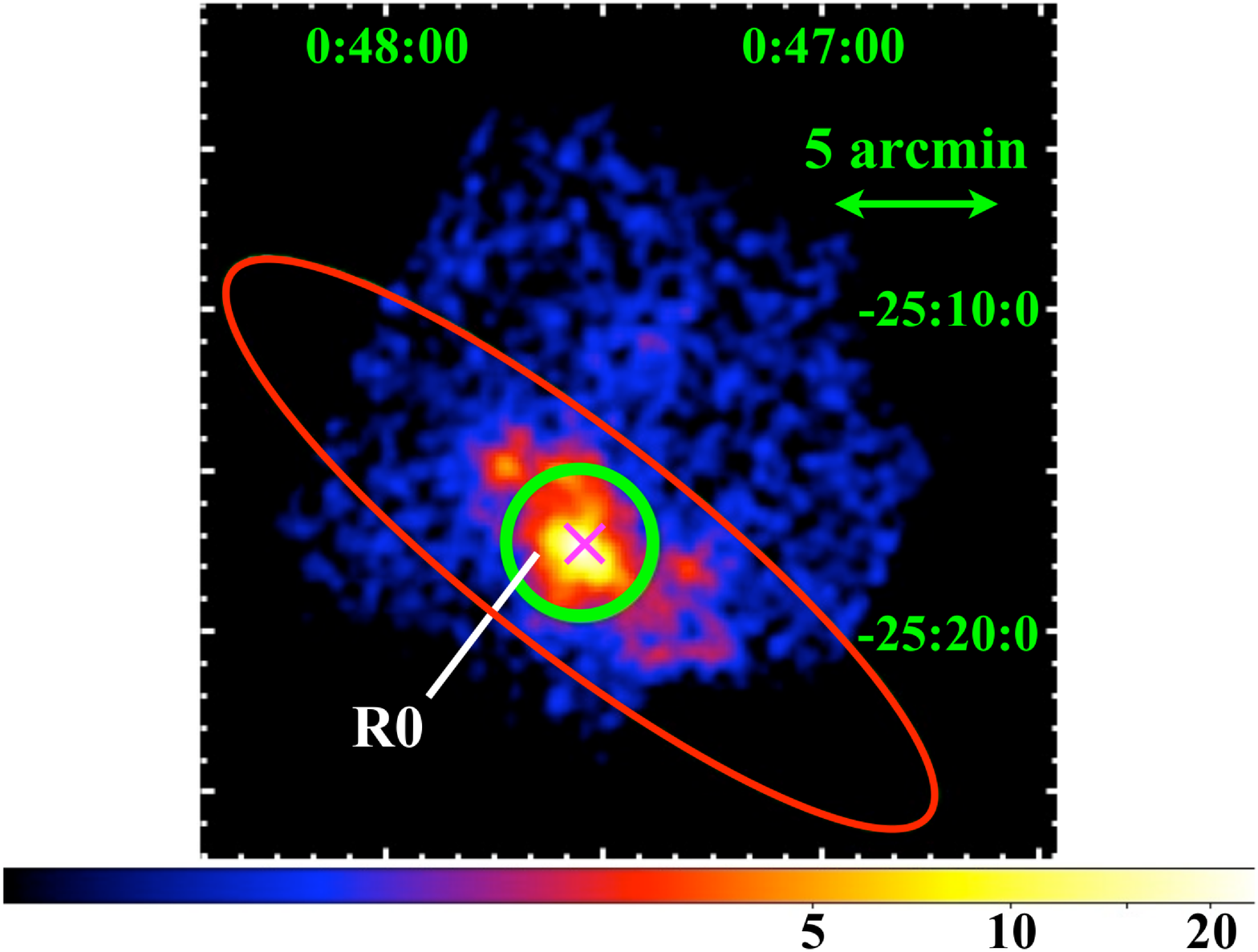} 
\includegraphics[width=5.4cm]{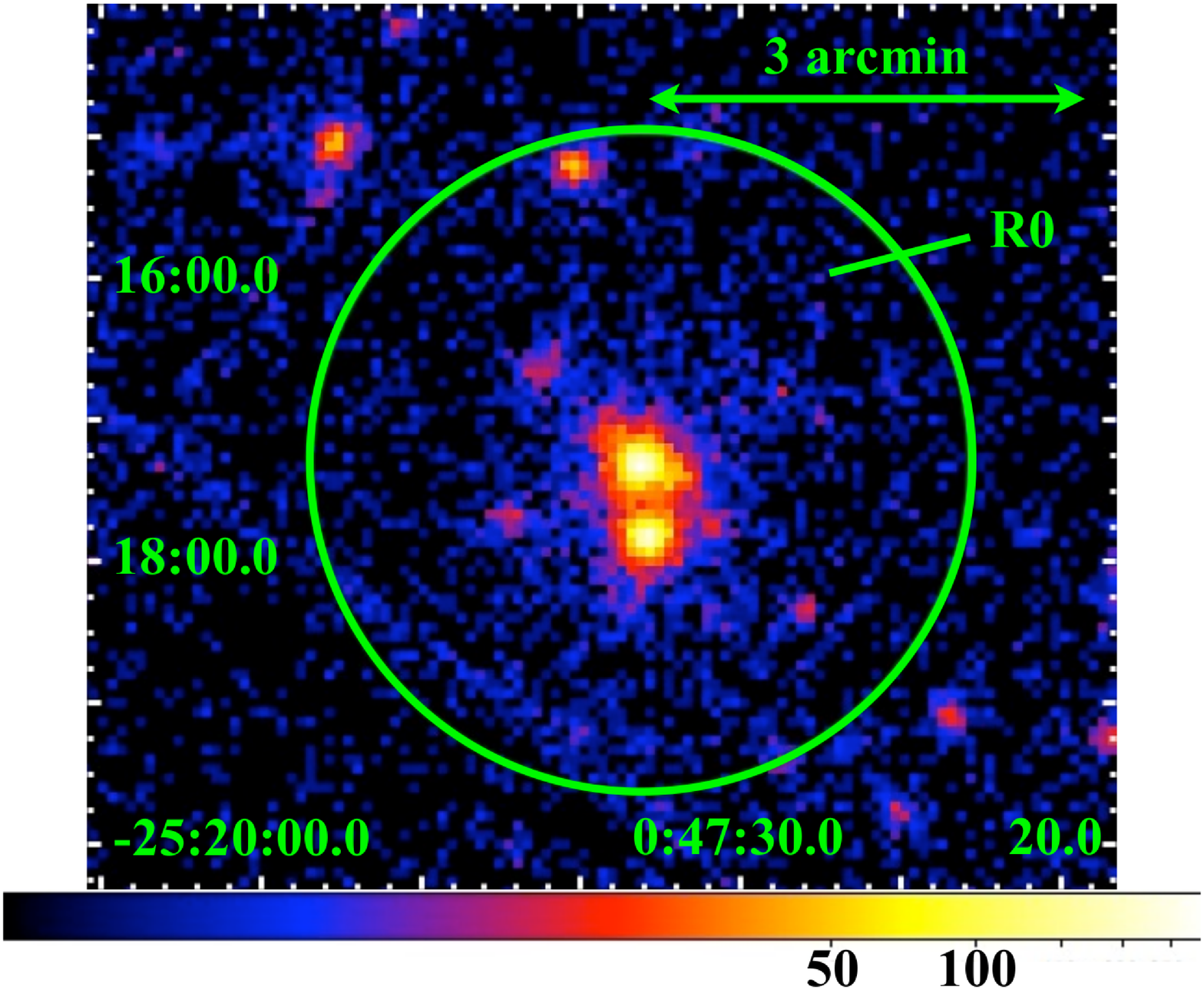}  
\includegraphics[width=5.4cm]{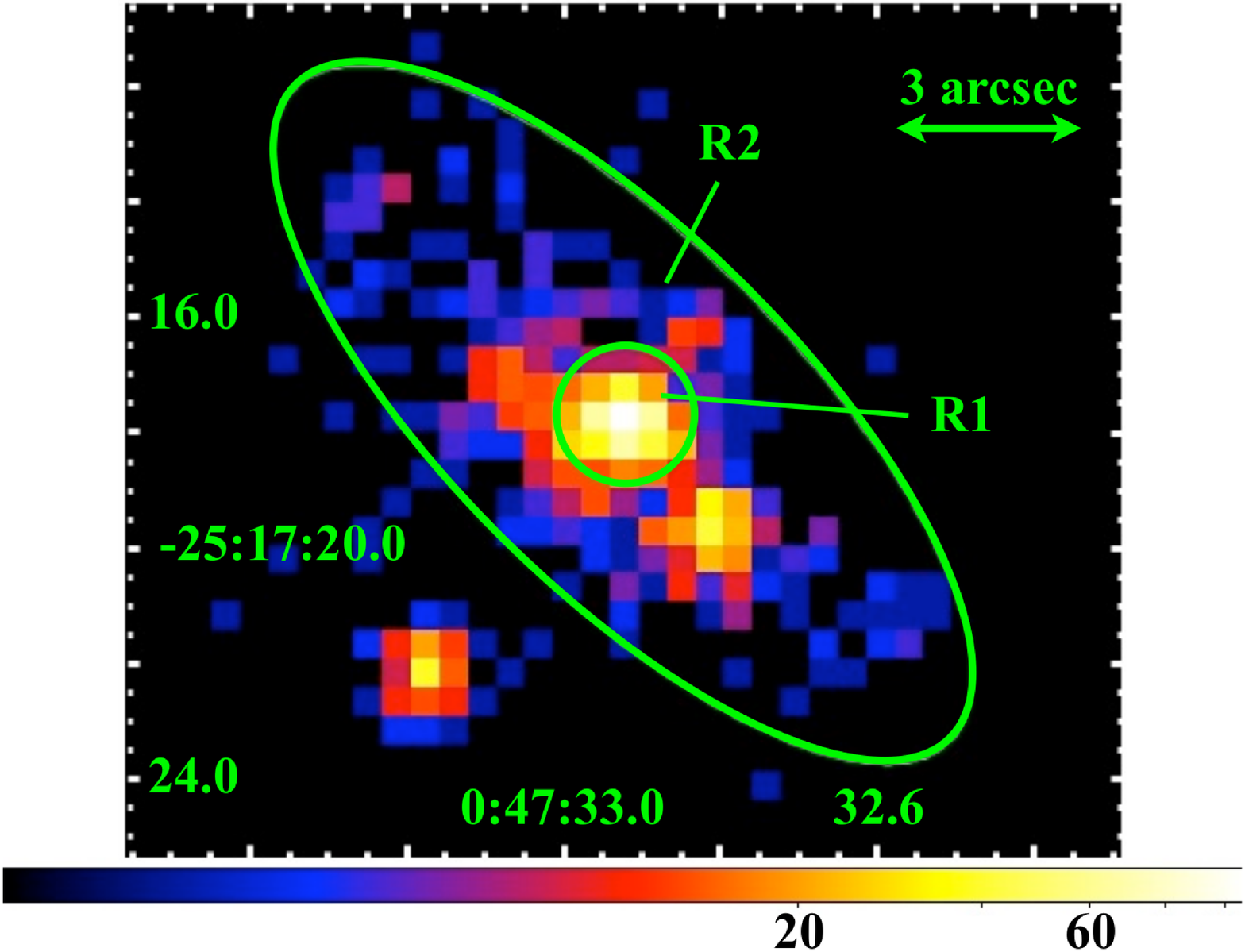}  
   \caption{Images taken by {\it Suzaku} XIS3 (left), 
   {\it XMM} PN (middle) and {\it Chandra} ACIS (right) in 4.0--10 keV.
  The red ellipse and magenta cross mark correspond to an optical disk and 
  the center of the {\it Chanda} image. Green circles and ellipses are regions used for extracting spectra.}
\label{fig:suzaku-xis3-chandra-acis-image-4-10keV}
\end{center}
   \end{figure*}

\objectname{NGC 253} was observed by {\it Suzaku} (seq:805018010) 
with an exposure time of 101 ks. 
We analyzed only XIS-FI (XIS0 and XIS3), which has a lower
background than XIS-BI (XIS1) at $>5~\mathrm{keV}$ to avoid a systematic uncertainty 
associated with a simultaneous fitting of different sensors.
The redistribution matrix files (RMFs) and 
ancillary response files (ARFs) were created with 
the xisrmfgen and xissimarfgen ftools \citep{arf}, respectively.
Since the bright region is within several tens of arcseconds,
which is much smaller than the {\it Suzaku } point spread function (PSF; $1'.8$ as a half-power diameter),
we assume the source to be a point-source when 
 creating the arf files.
The accumulation of 
dark Earth observations was used to estimate the non X-ray background
(NXB) spectra,
via the xisnxbgen ftool \citep{xisnxbgen}.
We subtracted NXB, the dominant background above 5 keV,  
but not the X-ray background, which was evaluated by an offset observation next to our field 
as to be less than 3 \% of the flux in R0.
The spectra and responses of XIS0 and XIS3
were merged using the ftools mathpha, marfrmf and addrmf.

We utilized PN data from two {\it XMM} observations (ObsID:0125960101 and 0152020101) 
with the effective exposure time of 31 and 54 ksec after removing flared time intervals  
which is over 0.35 counts/sec above 10 keV.
RMS and ARF files were made by the SAS rmfgen and arfgen tools.
We extracted the background from a source-free region.

The object was also  observed by {\it Chandra} three times 
with the net exposure time of 14 ks (seq:600114), 44 ks (seq:600093) 
and 83 ks (seq:600305), respectively. 
We analyzed the level 2 {\it Chandra} ACIS-S event data.
We used the merged data of three observations without any good-time-interval filtering 
because the background count rate is always $<$6.5 \% of the source in 5.0--8.0 keV.
RMFs and ARFs were created using the CIAO mkacisrmf and mkwarf tools. 
We extracted the background from a source-free region.

\section{Analysis \& Results}
\label{SEC:analysis-results}

\subsection{Neutral, He-like and H-like Fe K lines with {\it Suzaku} and {\it XMM}}

Spectra taken with {\it Suzaku} were
extracted from a circular region with a radius of $2'.3$,
indicated as R0 in Figure~\ref{fig:suzaku-xis3-chandra-acis-image-4-10keV} left.
The center of R0 corresponds to the peak position of the {\it Suzaku} 4.0--10.0 keV image.   
The size of R0 was optimized to realize the largest number of photons avoiding other disk-origin components 
and anomaly columns in the XIS0 detector below the nuclear region.

\begin{figure*}[htb!]
\begin{center}
  \includegraphics[width=5.cm]{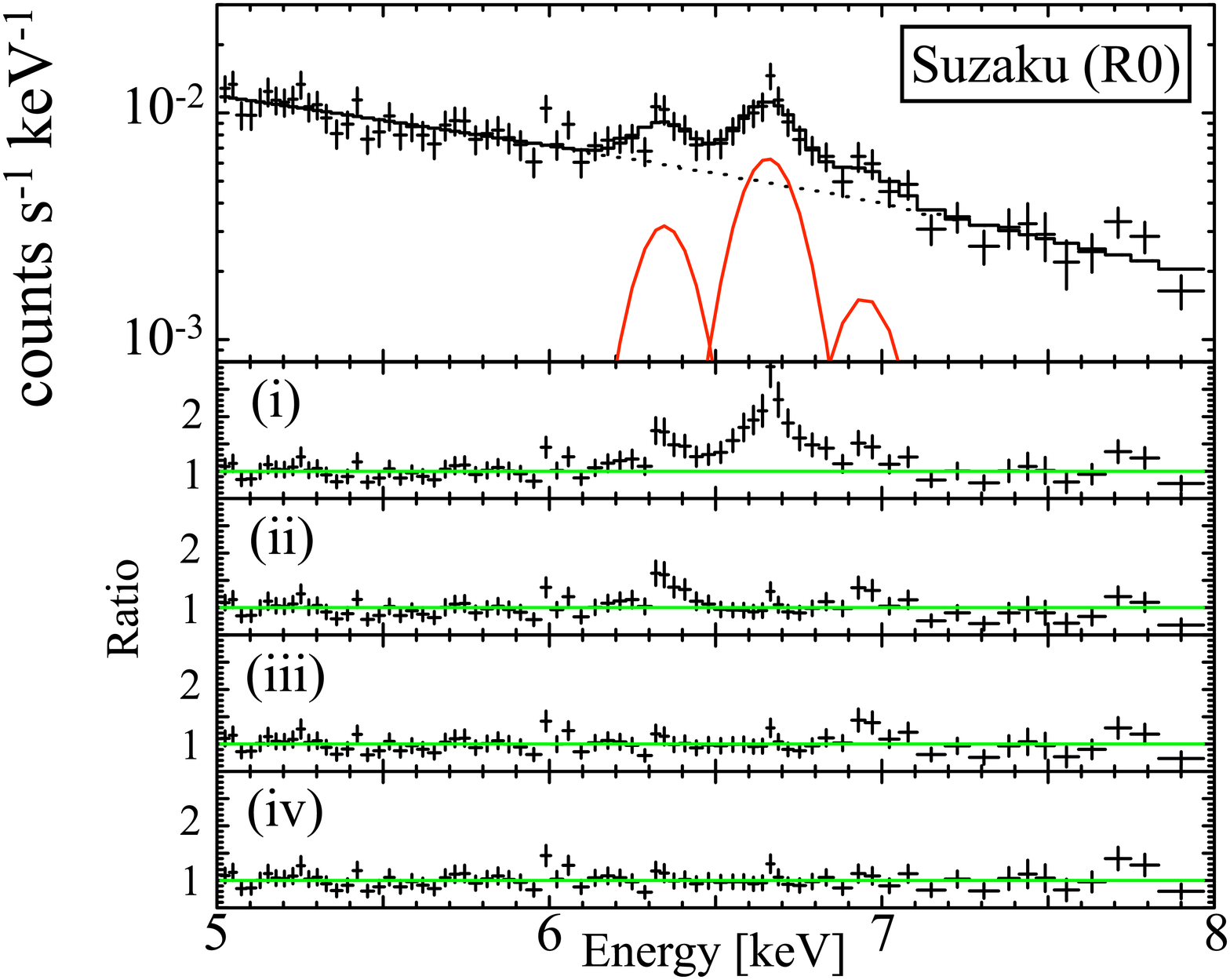} \hspace{0.2cm}
  \includegraphics[width=5.cm]{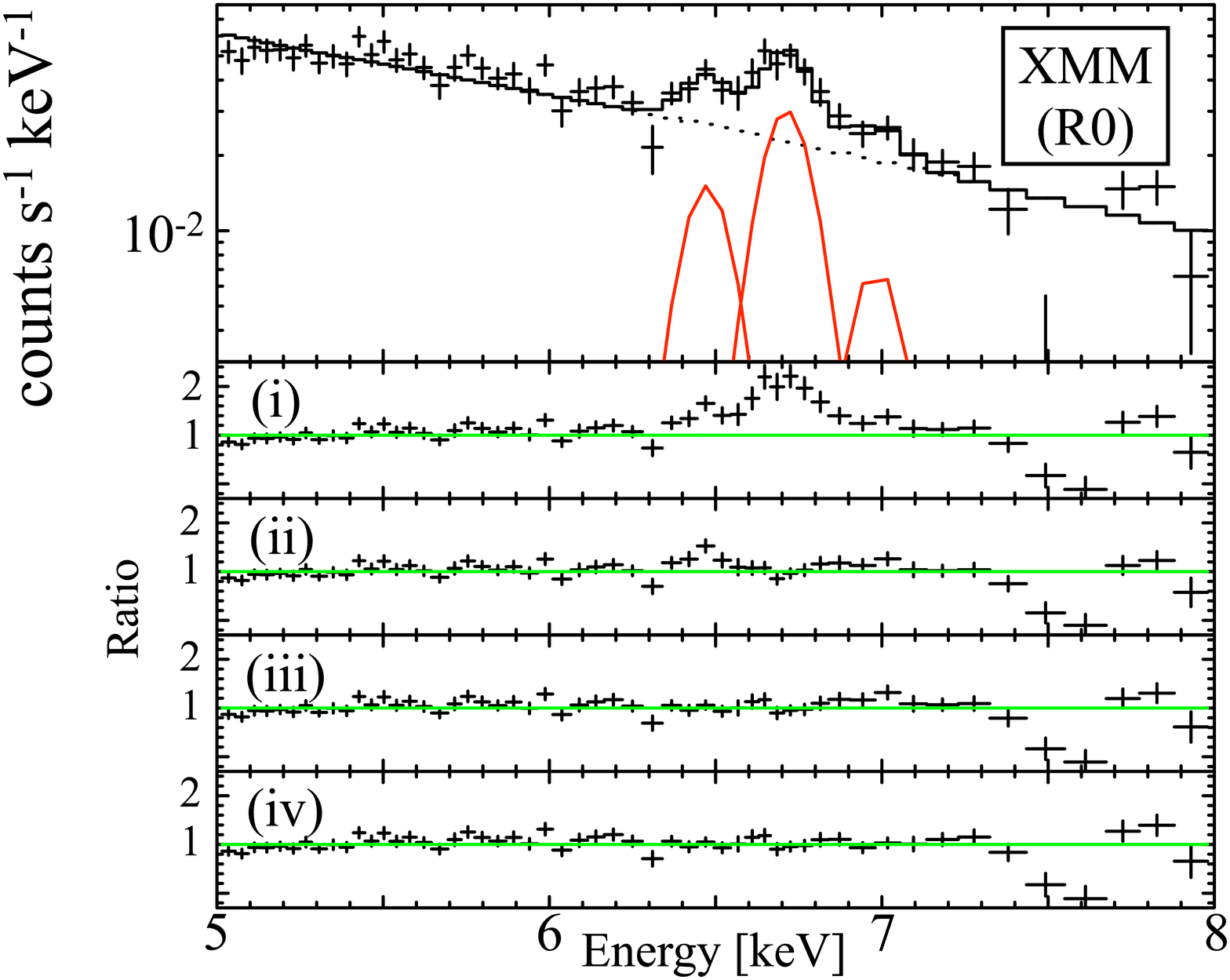} \hspace{0.2cm}
\includegraphics[width=5.cm]{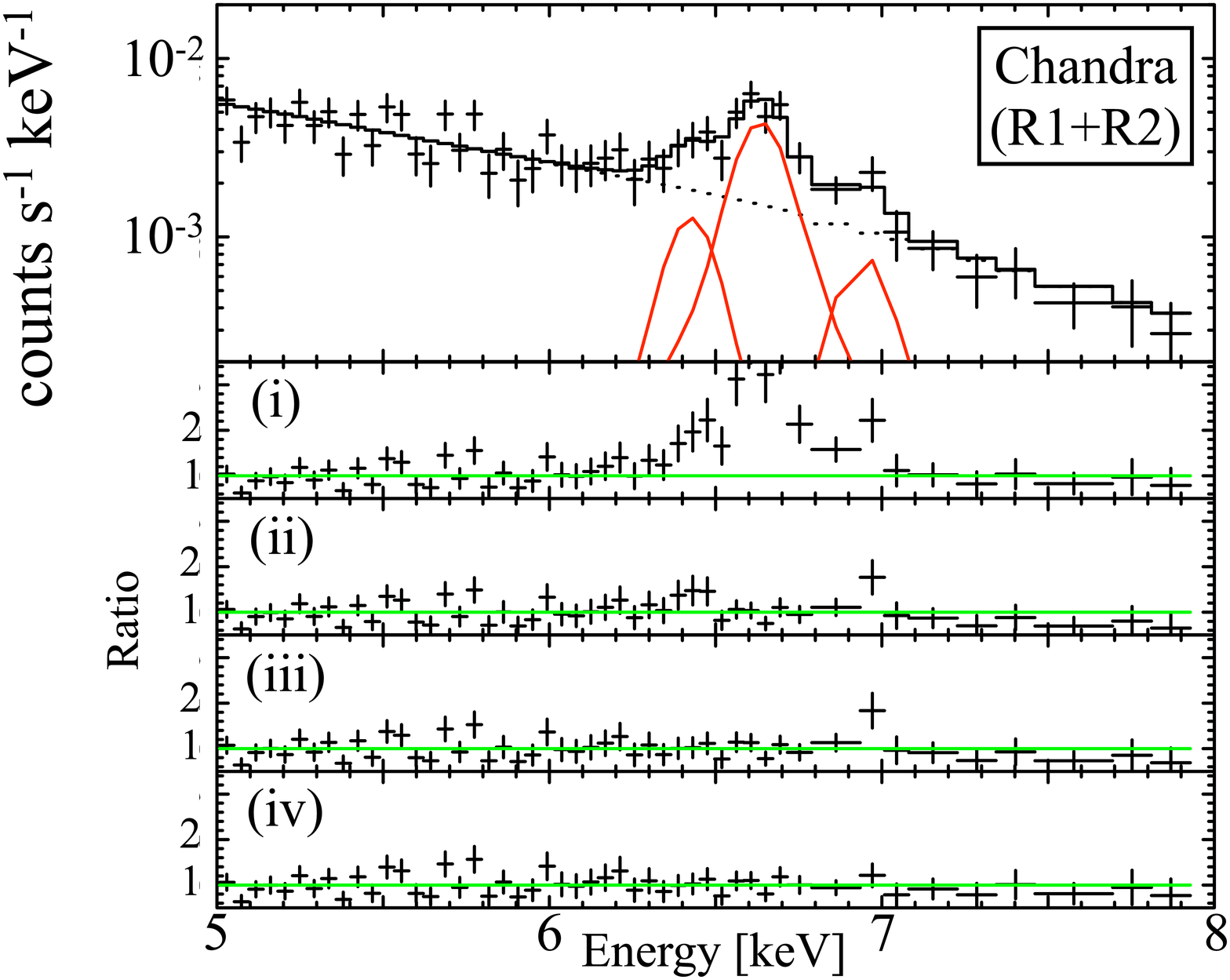}  
   \caption{Spectra with residuals in 5.0--8.0 keV of {\it Suzaku} FI (left), 
{\it XMM} PN (middle) from R0 and {\it Chandra} ACIS from (R1 + R2). 
The regions are defined in Figure \ref{fig:suzaku-xis3-chandra-acis-image-4-10keV}. 
Each residual panel indicates that to the model of
{\it power-law}, (ii) {\it power-law + gaussian}, 
(iii) {\it power-law + 2gaussian} and (iv) {\it power-law + 3gaussian}. 
The power-law continuum  was determined by excluding 6.3--7.1 keV 
to avoid the contamination from Fe K emission lines.}
\label{fig:suzaku-fi-chandra-acis-fe-k-complex-line-5-8keV}
\end{center}
   \end{figure*}

As shown in 
Figure~\ref{fig:suzaku-fi-chandra-acis-fe-k-complex-line-5-8keV} left,
complex Fe K line structure
was detected from the {\it Suzaku} spectrum as large residuals to
a power-law continuum model.
The continuum was
determined in 5.0--8.0 keV but excluding 6.3--7.1 keV 
to avoid Fe K emission lines.
We added Gaussian emission line models one by one.
The line width was fixed to 0. 
Fitting was significantly improved by each step.
The best-fit line center, equivalent width (EW), and flux,
along with their errors are summarized in Table~\ref{table:complex-fe-k-line-center-sb}.
We considered only the error of line flux in estimating  the error of EW.
A systematic error in flux
due to PSF modeling is bracketed by comparing two arfs: one for a point source and one for a uniformly extended source. 
The difference, $\sim$10 $\%$, was smaller than the statistical error. 
{\it Suzaku} XIS achieved an energy scale accuracy better than 6 eV at 6 keV
 \citep{suzaku-xis-koyama}, which means that 
the systematic error on the energy center is much smaller than the statistical error.
The 3 $\sigma$ acceptable range of  the energy center of the lowest-energy line 
from our {\it Suzaku} observation is 6.29-6.43 keV, 
which includes  \ion{Fe}{1} to \ion{Fe}{16} 
 \citep{line-center-house1969}.
The most natural explanation
of the line is the neutral fluorescent line.  We therefore interpreted the line as
\ion{Fe}{1}, although
\citet{chandra-nuclear-agn-suggestion} indicated that it could be emission 
from low ionization-state ions (\ion{Fe}{4}-\ion{Fe}{15}).

We employed the Monte Carlo procedure described in \citet{whim-ovii-absorption-sculptor-2009} 
to calculate a significance of these lines.
In this process, we evaluated the reduction in the C-statistic 
(29.9 for \ion{Fe}{1} and 11.4 for \ion{Fe}{26})
after adding a line with the  energy center range restricted to 6.3--6.5 keV for the \ion{Fe}{1} line 
and 6.87--7.07 keV for the \ion{Fe}{26} line.
The two lines were added individually.
We conducted 10000 Monte Carlo simulations without  \ion{Fe}{1} and \ion{Fe}{26} lines
with the same exposure and binning as the observation.
The fraction of simulation that improved the C-statistic by 29.9 and 11.4 are
0 and $1.1\times 10^{-3}$, respectively.
Therefore the significance of the detection of
\ion{Fe}{1} and \ion{Fe}{26} are $>$99.99~\% and 99.89~\%, respectively.
This is the first clear detection of the \ion{Fe}{1} and \ion{Fe}{26}
lines in \objectname{NGC 253}.

We also extracted a spectrum from R0 by {\it XMM} as a complementary analysis. 
We found the same complex Fe K lines as shown in Figure \ref{fig:suzaku-fi-chandra-acis-fe-k-complex-line-5-8keV} middle and 
we confirmed that resultant parameters are almost consistent with those of {\it Suzaku} which are summarized in Table \ref{table:complex-fe-k-line-center-sb}.
\renewcommand{\baselinestretch}{1.5}\selectfont
\begin{table}[htbp]
\begin{center}
\caption{Energy and flux of complex Fe K lines.}
\label{table:complex-fe-k-line-center-sb}
\scriptsize
\begin{tabular}{ccccc} \hline\hline
                                                          Region &
 & neutral Fe                              & Fe~XXV        & Fe~XXVI
 \\   \hline
\multicolumn{5}{c}{\it Suzaku} \\ \hline

&Energy (keV) $^{\ast}$                                                   & 6.36$\pm$0.03                     & 6.67$\pm$0.02                     & 6.96$^{+0.05}_{-0.08}$   \\  
  R0                            &Flux $^{\dagger}$                        & 3.8$^{+1.1}_{-1.0}$              & 8.4$^{+1.2}_{-1.1} $           & 2.3$^{+0.9}_{-1.2}$   \\ 
                                   &EW (eV)                                          & 84$^{+24}_{-22} $               & 220$^{+31}_{-29} $             & 60$^{+23}_{-31} $     \\ \hline
\multicolumn{5}{c}{\it XMM} \\ \hline

&Energy (keV) $^{\ast}$                                                   & 6.49$\pm$0.02                     & 6.72$\pm$0.01                     & 6.99$\pm$0.03   \\  
  R0                            &Flux $^{\dagger}$                        & 3.8$^{+1.0}_{-0.8}$              & 8.0$\pm$0.9                         & 2.1$^{+0.8}_{-0.9}$   \\ 
                                   &EW (eV)                                          & 77$^{+20}_{-16} $               & 189$\pm$21                         & 52$^{+20}_{-22} $     \\ \hline
                                   \multicolumn{5}{c}{\it Chandra } \\ \hline
&Energy (keV) $^{\ast}$                                                      & 6.40 (fix)                               & 6.66$^{+0.05}_{-0.06}$       & 6.97 (fix)   \\  
 R1                             & Flux $^{\dagger}$                        & $<$0.9                                    & 2.1$\pm$1.2                          & $<$1.1           \\
                                   &  EW (eV)                                        & $<$33                                     & 253$\pm$145                        &  $<$91                  \\ \hline
&Energy (keV) $^{\ast}$                                                    & 6.45$^{+0.19}_{-0.09}$      & 6.64$^{+0.04}_{-0.01}$        & 6.97 (fix)   \\  
 R2                              & Flux $^{\dagger}$                       & 1.2$\pm$1.1                         & 4.8$^{+1.4}_{-1.1}$               & $<$0.8   \\
                                    &  EW (eV)                                        &  86$\pm$79                         & 617$^{+180}_{-141}$           & $<$59     \\ \hline
& Energy (keV) $^{\ast}$                                                    & 6.43$^{+0.20}_{-0.08}$      & 6.64$^{+0.02}_{-0.03}$       & 6.95$^{+0.13}_{-0.12}$   \\  
  R1+R2                    & Flux $^{\dagger}$                        & 1.7$^{+1.1}_{-1.3}$             & 7.1$^{+1.5}_{-0.9}$               & 1.7$\pm$1.2s   \\ 
                                   &  EW (eV)                                        & 75$^{+49}_{-57} $               & 427$^{+90}_{-54}$                & 82$\pm$58   \\ \hline
& Energy (keV) $^{\ast}$                                                    & 6.46$^{+0.20}_{-0.07}$      & 6.65$\pm$0.04                      & 6.97 (fix)   \\  
P1                              & Flux $^{\dagger}$                       & 0.8$^{+0.7}_{-0.4}$              & 1.9$^{+1.0}_{-0.6}$               & $<$0.9   \\
                                   &  EW (eV)                                        & 176$^{+154}_{-88}$         & 662$^{+348}_{-209}$           & $<$73    \\ \hline
\end{tabular}
\end{center}
\begin{flushleft} 
\footnotesize{$^{\ast}$ Energy center at the rest frame.\\
$^{\dagger}$ Photon flux in the unit of 10$^{-6}$ photons s$^{-1}$ cm$^{-2}$.}
\end{flushleft}
\end{table}
\renewcommand{\baselinestretch}{1.0}\selectfont
\subsection{Spectral Analysis with Chandra}

We utilized {\it Chandra}, characterized 
by a sub-arcsecond angular resolution, for determining 
the emission region of the Fe K lines.
From a {\it Chandra} image, we see that
most of the emission above 4~keV is concentrated in $\sim 8''$ area 
(Figure \ref{fig:suzaku-xis3-chandra-acis-image-4-10keV} right).
We divided the compact area into two regions, R1 and R2, as indicated 
in Figure \ref{fig:suzaku-xis3-chandra-acis-image-4-10keV} right to examine 
whether the emission lines are associated with the AGN. 
R1 is a circlular region of a radius of $1''.2$
whose center is the peak of 4--10 keV surface brightness.
The buried AGN is located in R1 \citep{ngc253-agn}.
R2 is an ellipse with the major and minor axes of 
$8''$ and $3''$, respectively, but excluding R1. 
First, we extracted the spectrum from the whole (R1 + R2) region.
The spectrum around the Fe K line complex is shown in
Figure \ref{fig:suzaku-fi-chandra-acis-fe-k-complex-line-5-8keV} right. 
We performed the same analysis as we did with the {\it Suzaku} data.
The results are summarized in Table \ref{table:complex-fe-k-line-center-sb}.
Flux of \ion{Fe}{25} and \ion{Fe}{26} are consistent 
among three observatories to within statistical errors, 
which suggests that these lines mostly originate from this compact region.
On the other hand, the flux of \ion{Fe}{1} observed with {\it Suzaku} and {\ XMM} is
twice larger than that with {\it Chandra}.
Hence, the neutral emission may be more extended than the ionized emission or variable.
The smaller EWs of the {\it Suzaku} and {\it XMM} observations are because of a higher
continuum due to the emission from the disk.

We also extracted the spectra from R1 and R2 
to evaluate the contribution from the buried AGN.
As is shown in Table \ref{table:complex-fe-k-line-center-sb},
\ion{Fe}{1} and \ion{Fe}{25} lines are observed in the R2 region,
while only an upper limit is obtained for \ion{Fe}{26}.
The contribution to \ion{Fe}{1} and \ion{Fe}{25} is even larger in
R2 than in R1.
Thus the AGN is not the dominant source of the two lines and 
these emission lines are distributed over the (R1+R2) regions. 

\subsection{Imaging analysis with Chandra}
\begin{figure}[htb!]
\begin{center}
\includegraphics[width=0.4\textwidth]{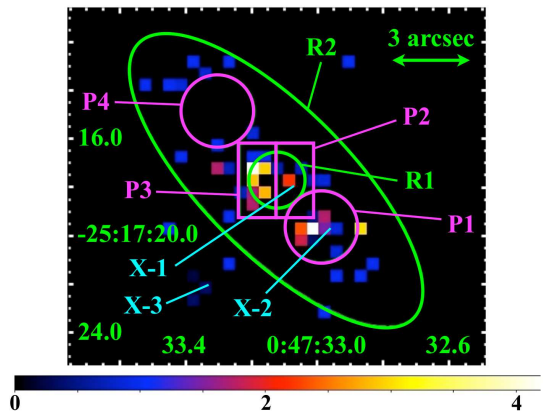}  
   \caption{Narrow band line images by {\it Chandra} for the \ion{Fe}{25} line (6.6--6.8 keV) 
   after subtracting a continuum contribution.
The color is in units of cts (141 ks)$^{-1}$ (1 pixel)$^{-1}$. The
 vignetting is not corrected for and the scale is linear.
Green and magenta regions indicate regions from which spectra were
extracted in this paper. Three X-ray sources (X-1, X-2, and X-3) found by \citet{ngc253-agn} are indicated by cyan.
}
   \label{fig:chandra-acis-line-maps}
 \end{center}
   \end{figure}

As described above, 
the observed Fe K complex is unlikely to be explained by the AGN alone.
In this section, we conducted imaging analysis 
to demonstrate the spacial distribution of this structure.
Figure \ref{fig:chandra-acis-line-maps} shows
the {\it Chandra} narrow band line flux image 
for the \ion{Fe}{25} (6.6--6.8~keV) line.
Contribution of the continuum was subtracted 
from the images in the following way.
We simply assumed that the continuum above 5 keV 
is expressed by a single power-law 
with the same photon index (2.0) in the whole region.
The contribution from the continuum in 6.6--6.8~keV is calculated
pixel by pixel using the photon index and the counts in 5.0--6.0~keV,
where no line feature exists.
The assumption of a uniform photon index is rather crude, 
because this can be applied for only the same spectrum properties.
We also confirmed that the following results did not change 
with different assumption of photon indices (1.0 and 4.0). 
In R1, a heavily absorbed non thermal component 
from the buried AGN instead of a high temperature plasma
dominates the hard band above 3 keV.
Despite poor statistics, the map suggests that
the line emission is not concentrated in R1, i.e.,
the location of the buried AGN.
In contrast, two bright regions of \ion{Fe}{25} line are found in P1 and P3 
defined in Figure \ref{fig:chandra-acis-line-maps}.

We also extracted the line flux image for the neutral Fe K line. 
However, the statistics are too poor to do any analysis. 
Hence we focus on the \ion{Fe}{25}.
To confirm the strong line emission in P1 and P3, 
we extracted spectra from four regions as indicated in Figure \ref{fig:chandra-acis-line-maps} 
denoted as P1, P2, P3 and P4.
P1 and P4 are circles with a radius of 1$''$.5 while P2 and P3 are rectangles of 1$''$.5 $\times$ 3$''$.
P1 and P3 include the peak position while 
P2 and P4 are selected from darker areas for comparisons.
P1 shows a strong line feature, 
while no line feature is found in P4.
The resulting line fluxes in P1 are shown in Table \ref{table:complex-fe-k-line-center-sb}.
$\sim$50 \% and 30 \% of the observed total \ion{Fe}{1} and \ion{Fe}{25} lines are emitted 
only from P1 as shown in Figure \ref{fig:chandra-acis-spectra} left.
P2 and P3 have the same continuum flux but the resulting line flux of P3 for \ion{Fe}{25} 
is twice larger than that of P2.

The coordinates of peak pixel centers in P1 and P3 are 
($\alpha_{2000}$, $\delta_{2000}$) = (00$^{{\rm h}}$47$^{{\rm m}}$33$^{{\rm s}}$.20, $-$25$^{\circ}$17$'$17$''$.2) and 
(00$^{{\rm h}}$47$^{{\rm m}}$33$^{{\rm s}}$.01, $-$25$^{\circ}$17$'$19$''$.7), respectively.
These are coincident with the peak positions of 1.3 mm continuum in the radio observation \citep{ngc253-radio-sakamoto} 
to within the position resolutions.
The 99 \% uncertainty of absolute X-ray position by {\it Chandra} is 0$''$.8, 
while the beam size of the radio observation is 1$''$.1$\times$1$''$.1. 
Thus, the \ion{Fe}{25} line shows a clear association with the two peaks of molecular clouds.
P1 is the most luminous in mid infrared 
and is suggested to be a super star cluster \citep{ngc253-radio-keto}. 
P3 is the brightest in the 1.3~mm radio continuum and
is featured by the water maser associated with star 
formation \citep[e.g.][]{ngc253-radio-henkel}.

\subsection{Wide-band Spectrum fitting of the P1 region}
\begin{figure*}[htb!]
\begin{center}
\includegraphics[width=8cm]{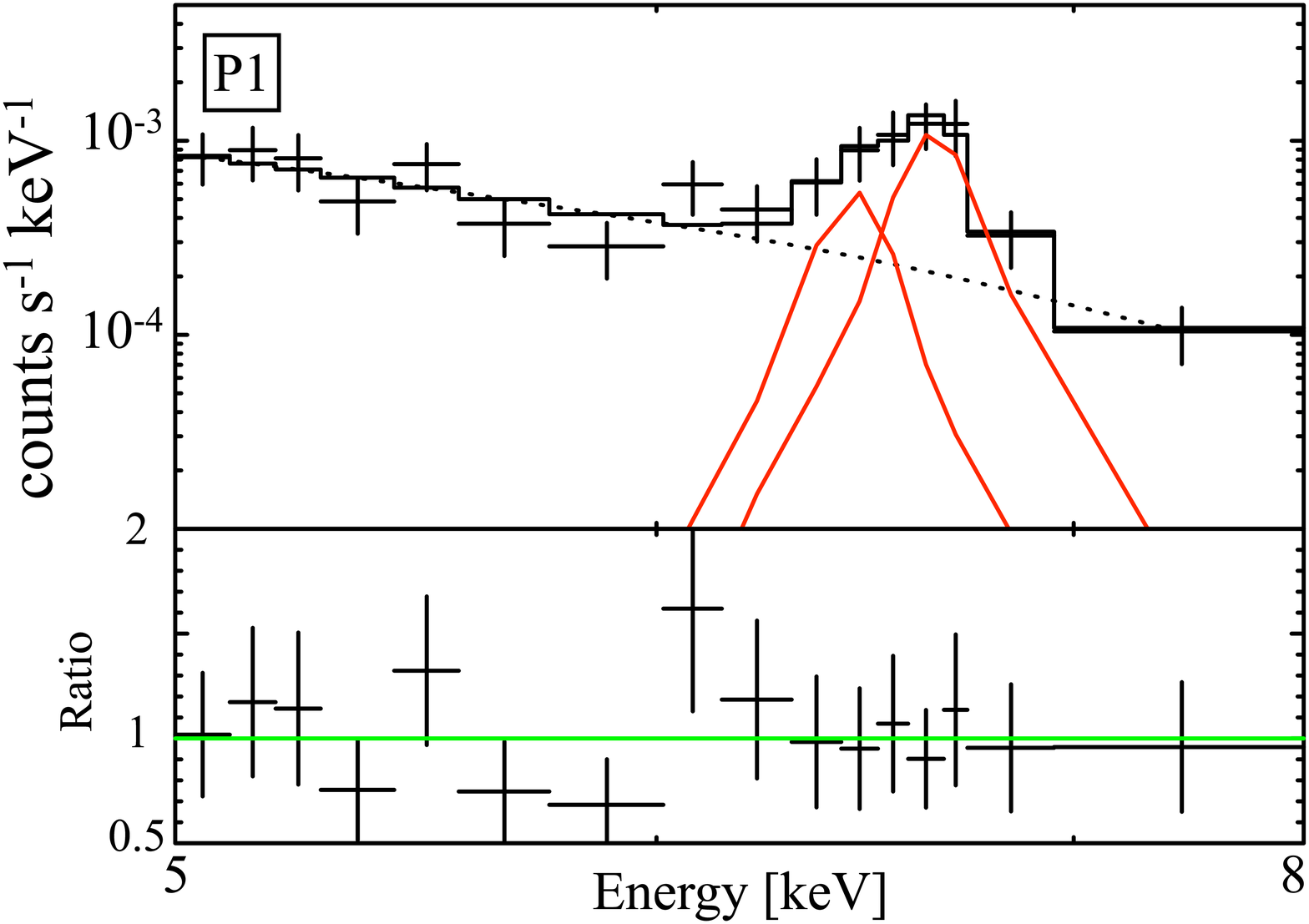}  
\includegraphics[width=8cm]{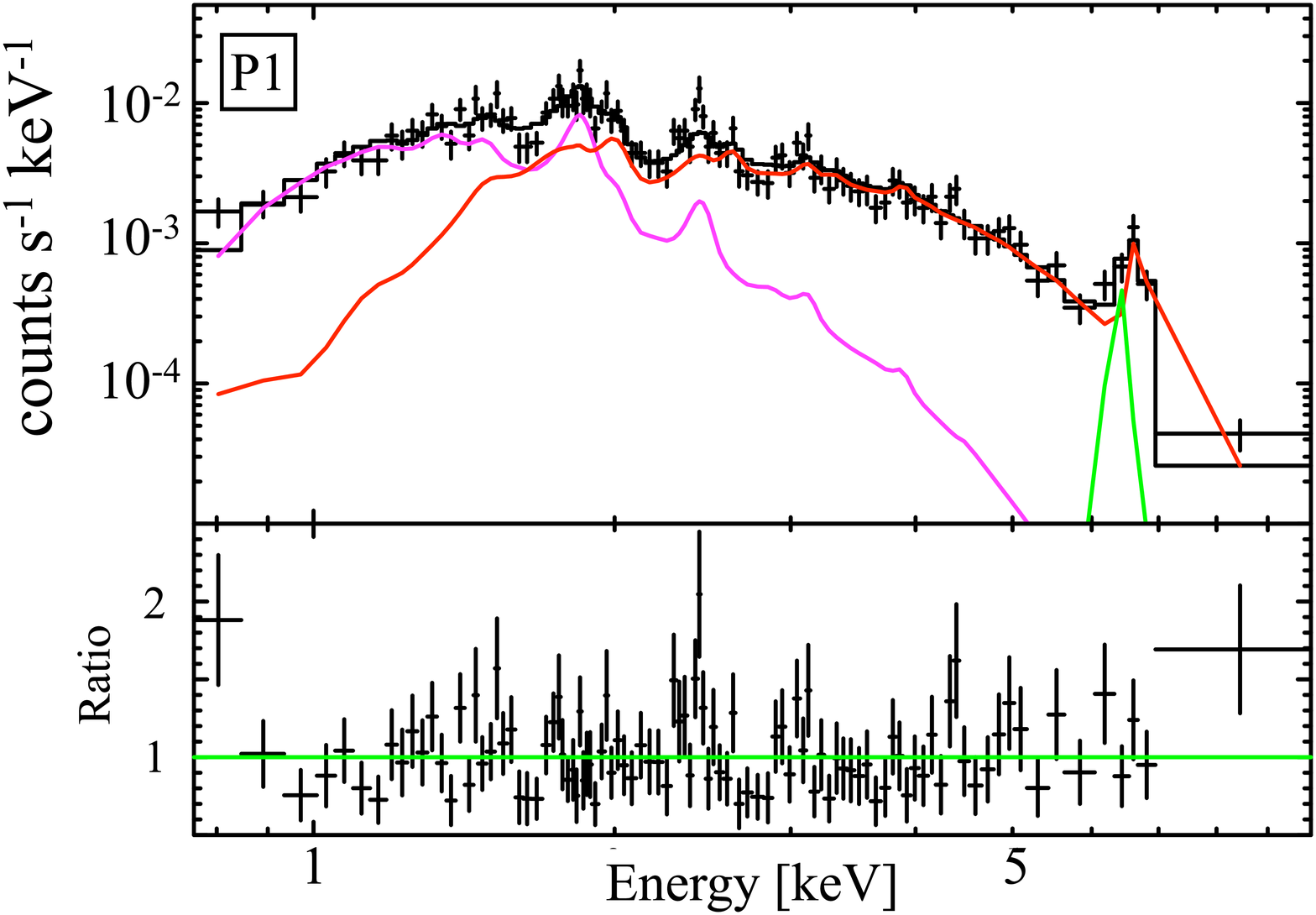}  
\end{center}
   \caption{{\it Chandra} ACIS-S spectra of P1. 
   Left: spectrum in 5.0--8.0 keV fitted with {\it power-law + 2gaussian} corresponding to \ion{Fe}{1} and \ion{Fe}{25} lines (red). 
   Right: wide-band spectrum in 0.8--10 keV. Magenta, red and green lines indicate thin thermal plasmas and the neutral Fe K line. 
\label{fig:chandra-acis-spectra}}
   \end{figure*}

To extract physical parameters 
we fitted the wide-band spectra (0.8--10~keV) 
of P1, where the highest line flux was confirmed. 
We adopted two-temperature thermal plasmas in collisionally ionization equilibrium 
plus a Gaussian whose centroid is 6.4 keV as the neutral Fe K line.
One plasma represents highly-ionized Si and S emission lines 
and the other makes up for continuum above 3 keV 
and the \ion{Fe}{25} emission line.
The spectrum with the best-fit model is plotted in
Figure \ref{fig:chandra-acis-spectra}.
The abundance is fixed at solar values.
The spectrum is well represented with the ${\rm C~statistic}/d.o.f~{\rm of}~132/95$.
The temperature of the higher-temperature plasma is 2.5$^{+0.5}_{-0.3}$ keV.
This plasma is responsible for the \ion{Fe}{25} and \ion{Fe}{26} lines.
The unabsorbed luminosity, thermal energy and total iron mass of
the higher-temperature plasma are calculated to be 
(8.6$\pm1.4)\times$10$^{38}$~$\mathrm{erg~  s^{-1}}$, 5.4$^{+0.5}_{-0.6}$$\times$10$^{52}$~$\mathrm{erg}$ and 
20$\pm2$ M$_{\odot}$, respectively.
The unabsorbed luminosities of the \ion{Fe}{1} and \ion{Fe}{25} lines in P1 are also 
calculated to be 1.2$^{+1.3}_{-0.9}~\times10^{37}$ and 2.7$^{+1.2}_{-1.5}~\times$10$^{37}$~$\mathrm{erg~  s^{-1}}$.
Note that P1 has a radius of 24 pc and the spherical volume of 5.8$\times$10$^4$ pc$^3$.

\section{Origin of the emission lines}

In this section, we discuss the origin of the neutral \ion{Fe}{1}, ionized \ion{Fe}{25} and \ion{Fe}{26} lines. 
This line structure likely extends to the R2 region, i.e., the central 8$''$ corresponding to $\sim$130 pc 
in the nuclear region of \objectname{NGC253}.
However, we focus on the very compact P1 region with a radius of 24 pc, 
which shows the most luminous line structure.
We calculate the contribution from
(i) point sources such as cataclysmic variables (CVs), active binaries and high-mass X-ray binaries (HMXBs), 
(ii) SNRs
and (iii) cold molecular clouds. 
Candidate (i) would produce all three lines, while
(ii) and (iii) would only be responsible for the ionized lines and 
the neutral lines, respectively.

\subsection{Point sources}

CVs and active binaries are bright point sources that
emit Fe lines with typical luminosities of 
$L_{2-10~{\rm keV}} \sim 10^{31}-10^{32}$ erg s$^{-1}$ and
$L_{2-10~{\rm keV}} < 10^{31}$ erg s$^{-1}$, respectively.
The luminosity in P1 corresponds to
$10^{7-8}$ sources, or a number density of
$\sim$100--1000 pc$^{-3}$.
We can also estimate the number density 
from line flux of CVs which are relatively luminous compared to active binaries.
The typical luminosity of the \ion{Fe}{1} line is on the order of 10$^{29-30}$ ~ $\mathrm{erg~  s^{-1}}$ 
\citep{magnetic-cv-fe-k-line-flux}, which leads to the 
large number 10$^{7-8}$ sources  and a number density of $\sim$500--5000 pc$^{-3}$.
The number density of CVs and active binaries are both
$10^{-5}-10^{-4}$ pc$^{-3}$ in the Galaxy
\cite[e.g.][]{cv-number-density4,
active-binary-number-density}.
Even at Galactic center, the expected number density of CVs is on the order of $10^{-2}-10^{-1}$ pc$^{-3}$, 
assuming that their number density can be scaled from the stellar density \citep{Galactic-center-cv-number-density}.
Therefore the contribution of these point sources seems low.
 HMXBs are also one of candidates with typical luminosity of 10$^{33-34}$ ~ $\mathrm{erg~  s^{-1}}$ 
 for \ion{Fe}{1} fluorescent line \citep{hmxb-fe-i-line-flux}.
This corresponds to 10$^{3-4}$ HMXBs or $10^{-2}-10^{-1}$ pc$^{-3}$ in P1.
Considering the number of HMXBs in the Galaxy ($\sim$100), contribution of HMXBs also seems low, 
unless extreme number of HMXBs and any other point sources such as black hole candidates 
are produced by star forming activity, or population of undetected obscured HMXBs 
\citep{supergiant-hmxb} is much larger. 

\subsection{SNRs}

Supernova remnants (SNRs) likely make the biggest contribution
as a diffuse plasma.
The luminosity and thermal energy of young or middle-aged 
 SNRs which contribute to highly-ionized Fe lines in LMC are
$10^{36-37}~ \mathrm{erg~  s^{-1}}$ and
$10^{51-52}~\mathrm{erg}$, respectively \citep{lmc-snrs}.
Both the luminosity and thermal energy of the higher-temperature plasma
of \objectname{NGC253} correspond to about
100--1000 and 10--100 young or middle-aged SNRs, respectively.
We also estimated the number of SNRs using 
the observed Fe mass (20$\pm$2 M$_{\odot}$) and 
Fe mass ejection per one Type II SN 
\cite[8.4$\times$10$^{-2}$ M$_{\odot}$;][]{snr-iwamoto}.
The resultant number of SNRs is about 240$\pm$20
which is consistent with the estimated value by the luminosity.
This suggests that the very dense massive star cluster in P1 exists,  
which is consistent with the results of the mid infrared observations.

\subsection{Molecular clouds}

Molecular clouds that are irradiated by surrounding X-ray sources would
emit a fluorescent \ion{Fe}{1} line.
We assume that the incident X-ray is photoelectrically absorbed by the molecular gas 
that subtends an angle $\Delta\Omega$ seen from the continuum source.
The line flux of the \ion{Fe}{1} K line can be estimated as 
\begin{equation}
I_{6.4~\mathrm{keV}} =
\left(\frac{\Delta\Omega}{4\pi}\right)\epsilon
\int^{\infty}_{7.1~\mathrm{keV}}dE
\int ds~
n_\mathrm{Fe}
\sigma_\mathrm{Fe}(E) F(E),
\end{equation}
where $\epsilon$, $n_\mathrm{Fe}$, $\sigma_\mathrm{Fe}(E)$ and $F(E)$ indicate 
the fluorescence yield of \ion{Fe}{1} K line ($\epsilon=0.34$), 
the density of the molecular gas, the photoelectric cross section, and the 
X-ray source intensity, respectively. 
Assuming a thermal bremsstrahlung spectrum for the continuum source, 
the X-ray source intensity, $F(E) \propto \exp(-E/k_BT)$, can be determined
in the 6.0--7.5~keV band.
The Fe column density is written as $\int n_\mathrm{Fe}ds = Z_\mathrm{Fe}N_H$,
where $Z_\mathrm{Fe}$ and $N_H$ represent 
the elemental abundance of iron to hydrogen and the column density of
the required molecular gas, respectively.
The photoelectric cross section of Fe above 7.1 keV is approximated 
as $\sigma_\mathrm{Fe}(E) = 3.8\times10^{-20}(E/7.1~\mathrm{keV})^{-2.58}$ cm$^2$
as described in \citet{photoelectric-cross-section-nobukawa2008}.
Substituting these values, we obtain 
 \begin{equation}
N_H= (1.0^{+0.9}_{-0.5})\times10^{24}\left(\frac{\Omega}{2\pi}\right)^{-1} ~\mathrm{cm^{-2}}.
 \end{equation}
The solid angle $\Omega=2\pi$ corresponds to the case that 
the X-ray sources are distributed entirely behind the molecular gas.  
This column density was consistent 
with the molecular clouds observed 
in the radio observation \citep{ngc253-radio-sakamoto} to within the statistical errors.
Thus, the observed \ion{Fe}{1} line flux is explained by the
emission from the \objectname{NGC253} nucleus and the 
dense molecular clouds in the nuclear region.

\section{Conclusions}

{\it Suzaku} and {\it XMM} observations revealed complex  Fe K line  structure in the nuclear region
of \objectname{NGC 253} and resolved it into \ion{Fe}{1}, \ion{Fe}{25}
and \ion{Fe}{26} lines.
{\it Chandra} demonstrated that the emission 
is concentrated mostly in the 
 $\sim$60 arcsec$^2$  nuclear region, 
 but is not localized at the position of a buried AGN. 
The \ion{Fe}{25} line shows an association with molecular clouds.
The complex Fe K line  can be explained by a combination of 10--1000 SNRs 
and a fluorescent line from molecular clouds ($N_H\sim 10^{24}~\mathrm{cm^{-2}}$).

\acknowledgments

I. M. is grateful to Kentaro Someya and Masahiro Tsujimoto for advices of the discussion on the origin. 
Part of this work was financially supported by the Ministry of Education, Culture, Sports, Science and 
Technology, Grant-in Aid for Scientific Research 10J07487, 20340068 and 22111513.


\begin{thebibliography}{26}
\expandafter\ifx\csname natexlab\endcsname\relax\def\natexlab#1{#1}\fi

\bibitem[{{Anders} \& {Grevesse}(1989)}]{solar-abundance-table-anders}
{Anders}, E., \& {Grevesse}, N. 1989, gca, 53, 197

\bibitem[{{Boller} {et~al.}(2003){Boller}, {Keil}, {Hasinger}, {Costantini},
  {Fujimoto}, {Anabuki}, {Lehmann}, \& {Gallo}}]{ngc6240-xmm}
{Boller}, T., {Keil}, R., {Hasinger}, G., {Costantini}, E., {Fujimoto}, R.,
  {Anabuki}, N., {Lehmann}, I., \& {Gallo}, L. 2003, \aap, 411, 63

\bibitem[{{Buote} {et~al.}(2009){Buote}, {Zappacosta}, {Fang}, {Humphrey},
  {Gastaldello}, \& {Tagliaferri}}]{whim-ovii-absorption-sculptor-2009}
{Buote}, D.~A., {Zappacosta}, L., {Fang}, T., {Humphrey}, P.~J., {Gastaldello},
  F., \& {Tagliaferri}, G. 2009, \apj, 695, 1351

\bibitem[{{Dalcanton} {et~al.}(2009){Dalcanton}, {Williams}, {Seth}, {Dolphin},
  {Holtzman}, {Rosema}, {Skillman}, {Cole}, {Girardi}, {Gogarten},
  {Karachentsev}, {Olsen}, {Weisz}, {Christensen}, {Freeman}, {Gilbert},
  {Gallart}, {Harris}, {Hodge}, {de Jong}, {Karachentseva}, {Mateo}, {Stetson},
  {Tavarez}, {Zaritsky}, {Governato}, \& {Quinn}}]{ngc253-distance}
{Dalcanton}, J.~J., {et~al.} 2009, \apjs, 183, 67

\bibitem[{{Ezuka} \& {Ishida}(1999)}]{magnetic-cv-fe-k-line-flux}
{Ezuka}, H., \& {Ishida}, M. 1999, \apjs, 120, 277

\bibitem[{{Favata} {et~al.}(1995){Favata}, {Micela}, \&
  {Sciortino}}]{active-binary-number-density}
{Favata}, F., {Micela}, G., \& {Sciortino}, S. 1995, \aap, 298, 482

\bibitem[{{Grimm} {et~al.}(2003){Grimm}, {Gilfanov}, \&
  {Sunyaev}}]{ngc253-sfrs}
{Grimm}, H.-J., {Gilfanov}, M., \& {Sunyaev}, R. 2003, \mnras, 339, 793

\bibitem[{{Henkel} {et~al.}(2004){Henkel}, {Tarchi}, {Menten}, \&
  {Peck}}]{ngc253-radio-henkel}
{Henkel}, C., {Tarchi}, A., {Menten}, K.~M., \& {Peck}, A.~B. 2004, \aap, 414,
  117

\bibitem[{{House}(1969)}]{line-center-house1969}
{House}, L.~L. 1969, \apjs, 18, 21

\bibitem[{{Hughes} {et~al.}(1998){Hughes}, {Hayashi}, \& {Koyama}}]{lmc-snrs}
{Hughes}, J.~P., {Hayashi}, I., \& {Koyama}, K. 1998, \apj, 505, 732

\bibitem[{{Ishisaki} {et~al.}(2007){Ishisaki}, {Maeda}, {Fujimoto}, {Ozaki},
  {Ebisawa}, {Takahashi}, {Ueda}, {Ogasaka}, {Ptak}, {Mukai}, {Hamaguchi},
  {Hirayama}, {Kotani}, {Kubo}, {Shibata}, {Ebara}, {Furuzawa}, {Iizuka},
  {Inoue}, {Mori}, {Okada}, {Yokoyama}, {Matsumoto}, {Nakajima}, {Yamaguchi},
  {Anabuki}, {Tawa}, {Nagai}, {Katsuda}, {Hayashida}, {Bamba}, {Miller},
  {Sato}, \& {Yamasaki}}]{arf}
{Ishisaki}, Y., {et~al.} 2007, \pasj, 59, 113

\bibitem[{{Iwamoto} {et~al.}(1999){Iwamoto}, {Brachwitz}, {Nomoto},
  {Kishimoto}, {Umeda}, {Hix}, \& {Thielemann}}]{snr-iwamoto}
{Iwamoto}, K., {Brachwitz}, F., {Nomoto}, K., {Kishimoto}, N., {Umeda}, H.,
  {Hix}, W.~R., \& {Thielemann}, F.-K. 1999, \apjs, 125, 439

\bibitem[{{Keto} {et~al.}(1999){Keto}, {Hora}, {Fazio}, {Hoffmann}, \&
  {Deutsch}}]{ngc253-radio-keto}
{Keto}, E., {Hora}, J.~L., {Fazio}, G.~G., {Hoffmann}, W., \& {Deutsch}, L.
  1999, \apj, 518, 183

\bibitem[{{Koyama} {et~al.}(2007){Koyama}, {Tsunemi}, {Dotani}, {Bautz},
  {Hayashida}, {Tsuru}, {Matsumoto}, {Ogawara}, {Ricker}, {Doty}, {Kissel},
  {Foster}, {Nakajima}, {Yamaguchi}, {Mori}, {Sakano}, {Hamaguchi},
  {Nishiuchi}, {Miyata}, {Torii}, {Namiki}, {Katsuda}, {Matsuura}, {Miyauchi},
  {Anabuki}, {Tawa}, {Ozaki}, {Murakami}, {Maeda}, {Ichikawa}, {Prigozhin},
  {Boughan}, {Lamarr}, {Miller}, {Burke}, {Gregory}, {Pillsbury}, {Bamba},
  {Hiraga}, {Senda}, {Katayama}, {Kitamoto}, {Tsujimoto}, {Kohmura}, {Tsuboi},
  \& {Awaki}}]{suzaku-xis-koyama}
{Koyama}, K., {et~al.} 2007, \pasj, 59, 23

\bibitem[{{M{\"u}ller-S{\'a}nchez} {et~al.}(2010){M{\"u}ller-S{\'a}nchez},
  {Gonz{\'a}lez-Mart{\'{\i}}n}, {Fern{\'a}ndez-Ontiveros}, {Acosta-Pulido}, \&
  {Prieto}}]{ngc253-agn}
{M{\"u}ller-S{\'a}nchez}, F., {Gonz{\'a}lez-Mart{\'{\i}}n}, O.,
  {Fern{\'a}ndez-Ontiveros}, J.~A., {Acosta-Pulido}, J.~A., \& {Prieto}, M.~A.
  2010, \apj, 716, 1166

\bibitem[{{Muno} {et~al.}(2003){Muno}, {Baganoff}, {Bautz}, {Brandt}, {Broos},
  {Feigelson}, {Garmire}, {Morris}, {Ricker}, \&
  {Townsley}}]{Galactic-center-cv-number-density}
{Muno}, M.~P., {et~al.} 2003, \apj, 589, 225

\bibitem[{{Nobukawa} {et~al.}(2008){Nobukawa}, {Tsuru}, {Takikawa}, {Hyodo},
  {Inui}, {Nakajima}, {Matsumoto}, {Koyama}, {Murakami}, \&
  {Yamauchi}}]{photoelectric-cross-section-nobukawa2008}
{Nobukawa}, M., {et~al.} 2008, \pasj, 60, 191

\bibitem[{{Pietsch} {et~al.}(2001){Pietsch}, {Roberts}, {Sako}, {Freyberg},
  {Read}, {Borozdin}, {Branduardi-Raymont}, {Cappi}, {Ehle}, {Ferrando},
  {Kahn}, {Ponman}, {Ptak}, {Shirey}, \& {Ward}}]{ngc253-nuclear-xmm}
{Pietsch}, W., {et~al.} 2001, \aap, 365, L174

\bibitem[{{Ranalli} {et~al.}(2008){Ranalli}, {Comastri}, {Origlia}, \&
  {Maiolino}}]{m82-xmm}
{Ranalli}, P., {Comastri}, A., {Origlia}, L., \& {Maiolino}, R. 2008, \mnras,
  386, 1464

\bibitem[{{Rogel} {et~al.}(2008){Rogel}, {Cohn}, \&
  {Lugger}}]{cv-number-density4}
{Rogel}, A.~B., {Cohn}, H.~N., \& {Lugger}, P.~M. 2008, \apj, 675, 373

\bibitem[{{Sakamoto} {et~al.}(2011){Sakamoto}, {Mao}, {Matsushita}, {Peck},
  {Sawada}, \& {Wiedner}}]{ngc253-radio-sakamoto}
{Sakamoto}, K., {Mao}, R.-Q., {Matsushita}, S., {Peck}, A.~B., {Sawada}, T., \&
  {Wiedner}, M.~C. 2011, \apj, 735, 19

\bibitem[{{Strickland} \& {Heckman}(2007)}]{m82-fe-k-complex}
{Strickland}, D.~K., \& {Heckman}, T.~M. 2007, \apj, 658, 258

\bibitem[{{Tawa} {et~al.}(2008){Tawa}, {Hayashida}, {Nagai}, {Nakamoto},
  {Tsunemi}, {Yamaguchi}, {Ishisaki}, {Miller}, {Mizuno}, {Dotani}, {Ozaki}, \&
  {Katayama}}]{xisnxbgen}
{Tawa}, N., {et~al.} 2008, \pasj, 60, 11

\bibitem[{{Torrej{\'o}n} {et~al.}(2010){Torrej{\'o}n}, {Schulz}, {Nowak}, \&
  {Kallman}}]{hmxb-fe-i-line-flux}
{Torrej{\'o}n}, J.~M., {Schulz}, N.~S., {Nowak}, M.~A., \& {Kallman}, T.~R.
  2010, \apj, 715, 947

\bibitem[{{Walter} {et~al.}(2006){Walter}, {Zurita Heras}, {Bassani},
  {Bazzano}, {Bodaghee}, {Dean}, {Dubath}, {Parmar}, {Renaud}, \&
  {Ubertini}}]{supergiant-hmxb}
{Walter}, R., {et~al.} 2006, \aap, 453, 133

\bibitem[{{Weaver} {et~al.}(2002){Weaver}, {Heckman}, {Strickland}, \&
  {Dahlem}}]{chandra-nuclear-agn-suggestion}
{Weaver}, K.~A., {Heckman}, T.~M., {Strickland}, D.~K., \& {Dahlem}, M. 2002,
  \apjl, 576, L19

\end{thebibliography}

\end{document}